 \documentclass[prd,aps,twocolumn,preprintnumbers, showpacs, nofootinbib,superscriptaddress,notitlepage]{revtex4-1}
 \usepackage{amssymb,amsthm,amsmath}
 \usepackage{textcomp}
 \usepackage{color}      
 \usepackage{slashed}    
 \usepackage{verbatim}
 \usepackage[normalem]{ulem} 
 
 \usepackage{soul}

 \usepackage{rotating}   
 \usepackage{multirow}   
 
\begin{document}
 	\title{ Complete determination of $SU(3)_F$ amplitudes and strong phase  in $\Lambda_c^+ \to \Xi^0 K^+$ } 
 	\author {
 		Chao-Qiang Geng}\email{cqgeng@ucas.ac.cn}
 	\affiliation{School of Fundamental Physics and Mathematical Sciences, Hangzhou Institute for Advanced Study, UCAS, Hangzhou 310024}
 	\author {
 		Xiao-Gang He}\email{hexg@sjtu.edu.cn}
 	\affiliation{Tsung-Dao Lee Institute,
 		Shanghai Jiao Tong University, Shanghai 200240, China} 
 	\affiliation{Department of Physics, National Taiwan University, Taipei, 10617
 	}
 	\author{ Xiang-Nan Jin}\email{xnjin@ucas.ac.cn}
 	\affiliation{School of Fundamental Physics and Mathematical Sciences, Hangzhou Institute for Advanced Study, UCAS, Hangzhou 310024}
 	\author {
 		Chia-Wei Liu}\email{
 		chiaweiliu@sjtu.edu.cn	}
 	\affiliation{Tsung-Dao Lee Institute,
 		Shanghai Jiao Tong University, Shanghai 200240, China} 
 	\author {
 		Chang Yang}\email{15201868391@sjtu.edu.cn}
 	\affiliation{Tsung-Dao Lee Institute,
 		Shanghai Jiao Tong University, Shanghai 200240, China} 
 	\date{\today}

 	\begin{abstract}
The BESIII collaboration has recently reported the first time measurement of the decay asymmetry $\alpha(\Lambda_c^+ \to \Xi^0 K^+) = 0.01 \pm 0.16(stat.) \pm 0.03(syst.)$ and also a sizable phase shift of  $\delta_P-\delta_S = -1.55  \pm 0.25$ or $1.59\pm 0.25$  between S- and P-wave amplitudes. This implies significant  strong phase shifts in the decay amplitudes. The strong phases indicate the existence of rescattering or loop effects, which are challenging to calculate due to non-perturbative effects. By employing the flavor \(SU(3)_F\) symmetry and applying the Körner-Pati-Woo theorem to reduce the number of parameters, we find that the current data already allow us  to obtain,  for the first time, model-independent decay amplitudes and their strong phases. The establishment of the existence of sizable strong phases opens a window for future investigations into CP violation. In our fit, a notable discrepancy emerges in the branching ratio of \(\Xi_c^0 \to \Xi^- \pi^+\). The direct relationship between \(\Gamma (\Lambda_c^+ \to \Lambda e^+\nu_e)\) and \(\Gamma (\Xi_c^0 \to \Xi^- e^+\nu_e)\), along with newly discovered \(SU(3)_F\) relations, collectively suggests an underestimation of \(\mathcal{B}(\Xi_c^0 \to \Xi^- \pi^+)\) in experimental findings.  
	\end{abstract}

 	\maketitle
 
 Recent results the BESIII collaboration have  reported  \( \alpha( \Lambda_c^+ \to {\Xi^0 K^+})= 0.01 \pm 0.16(stat.) \pm 0.03(syst.) \)~\cite{BESIII10}. This supplements the previously established \( {\cal B}(\Lambda_c^+ \to \Xi^0 K^+) = (0.55 \pm 0.07) \% \)~\cite{BESIII3}, highlighting the importance of this channel in deepening our understanding of baryon decays.    Moreover,
 BESIII data also indicates a non-zero $\beta(\Lambda_c^+ \to {\Xi^0 K^+})$ to be negative, implying  a strong phase shift between the S- and P-waves of $ \delta_P - \delta_S   =  - 1.55 \pm 0.25$ or $1.59 \pm 0.25$~\cite{BESIII10,idk}.  
These strong phases can be induced by rescattering processes and loop effects, where the intermediate particles are on-shell. This is an important feature originating from quantum theory and   an essential ingredient   for observing  CP violation in particle and anti-particle decays. In two-body baryonic decays, the strong phase shifts  manifest their effects in the Lee-Yang parameters \((\alpha, \beta)\)~\cite{Lee:1957qs}, which have  played significant roles in understanding weak interaction.

Theoretically, first principle calculations for these decay amplitudes are extremely difficult due the low energy scale involved where non-pertrurbative QCD effects become important.  Determinations for such decays need  to wait for a full lattice calculation. In the meantime, analyses of  low energy physics have proven to be useful~\cite{Gell-Mann:1962yej} with the help of the flavor $SU(3)_F$ symmetry to a good approximation~\cite{He:2018joe,Savage}. This flavor symmetry reduces the number of amplitudes by relating some of them together. When enough data are accumulated, it is possible to determine the decay amplitudes in a model-independent way   and make  testable predictions.

Efforts have been made in this direction recently. Previous  studies based on the   flavor $SU(3)_F$ symmetry predicted a large value close to one for $\alpha(\Lambda_c^+ \to \Xi^0 K^+)$  assuming real decay amplitudes~\cite{Asymmetries,Huang:2021aqu,Zhong:2022exp,Xing:2023dni} which also lead to zero strong phase shifts. The reasons for assuming real decay amplitudes were that there were no hints of strong phase shifts and also not enough data points to obtain useful information. The new data from BESIII now show the needs of having non-zero strong phase shifts, calling for a new theoretical understanding. In this work,  we show  that the decay amplitudes and their strong phases for two body weak decays of anti-triplet charmed baryons can be completely
determined from available data by applying the  K\"orner-Pati-Woo~(KPW) theorem to further reduce the number of   parameters~\cite{Korner:1970xq,Groote:2021pxt} and therefore explain  the measured non-zero strong phase. 
We achieved for the first time a model-independent   determination of two body decays of charmed anti-triplet baryon.

 \( \Lambda_c^+ \to \Xi^0 K^+ \) is one of the weak decays of a charmed anti-triplet baryon $(T_{c \bar 3})$
 to
 an octet charmless baryon $({\bf B} ) $ and 
  a nonet pseudoscalar $(P)$. 
 Their $SU(3)_F$ representations are given by 
 \begin{equation}
 	T_{c \bar 3}  =  (\Xi^0_c,\; - \Xi^+_c,\;\Lambda^+_c) \,, 
 \end{equation}
and 
\begin{widetext}
 \begin{equation}\small
{\bf B}
= 
\left(\begin{array}{ccc}
	\frac{1}{\sqrt{6}} \Lambda+\frac{1}{\sqrt{2}} \Sigma^0 & \Sigma^{+} & p \\
	\Sigma^{-} & \frac{1}{\sqrt{6}} \Lambda-\frac{1}{\sqrt{2}} \Sigma^0 & n \\
	\Xi^{-} & \Xi^0 & -\sqrt{\frac{2}{3}} \Lambda
\end{array}\right)\,,  ~~~
 P=  
\left(\begin{array}{ccc}
	\frac{1}{\sqrt{2}}\left(\pi^0+c_\phi \eta+s_\phi \eta^{\prime}\right) & \pi^{+} & K^{+} \\
	\pi^{-} & \frac{1}{\sqrt{2}}\left(-\pi^0+c_\phi \eta+s_\phi \eta^{\prime}\right) & K^0 \\
	K^{-} & \overline{K}^0 & -s_\phi \eta+c_\phi \eta^{\prime}
\end{array}\right)\,, 
 \end{equation}
\end{widetext}
where $(c_\phi, s_\phi) = (\cos \phi , \sin \phi)$ with $\phi = 39.3^\circ $~\cite{phin} the mixing angle. 

 

The decay amplitude for an initial baryon \( {\bf B}_i \) to a final baryon \( {\bf B}_f \) and a meson $P$,  can be written as:
\begin{equation}
	{\cal M} = \langle {\bf B}_f P | {\cal H}_{\text{eff}} | {\bf B}_i \rangle = 
	i \overline{u}_f \left(
	F - G\gamma_5
	\right)u_i\,,
\end{equation}
where \( u_{i(f)} \) denotes the Dirac spinor for the initial (final) baryon and  \( F \) (\( G \)) indicates a generic amplitude which violates (conserves) parity, associated with the S (P) partial wave. The values for \( F \) (\( G \)) depends on processes.
The decay width, \( \Gamma \), and the 
other  decay observables are given by:
\begin{align}
	\Gamma &= \frac{p_f}{8\pi}\frac{(M_i+M_f)^2-M_P^2}{M_i^2} \left( |F|^2+ 
\kappa^2 |G|^2\right), \nonumber \\
	\alpha &= \frac{2\kappa \text{Re}(F^* G )}{|F|^2+\kappa^2|G|^2}, \qquad \beta = 
	\frac{2\kappa \text{Im}(F^*G )}{|F|^2+\kappa^2|G|^2},\qquad \nonumber \\
	\gamma  &=  \frac{|F|^2-\kappa^2|G|^2}{|F|^2+\kappa^2|G|^2}, \qquad 
 F^* G  =  |F^*G| e^{i(\delta_P - \delta_S) }    
	\end{align}
where  \( M_{i,f} \) and \( M_P \) are the respective masses of \( {\bf B}_{i,f} \) and \( P \), 
$\kappa = 
p_f/(E_f+M_f)$, and 
$p_f( E_f )$  is the 3-momentum (energy) of \( {\bf B}_f \) in the rest frame of \( {\bf B}_i \).

The effective Hamiltonian inducing a charmed anti-triplet baryon weak decay is given by
	\begin{eqnarray}\label{3}
		{\cal H}_{eff} &=&
		\frac{G_F}{\sqrt{2} }
		\left( 
		c_+{\cal H}({\bf 15})^{ ij }_k
		+ 
		c_-{\cal H}(\overline{{\bf 6}})_{ l k  }\epsilon^{lij} 
		\right) 
	\nonumber\\
	&& (\overline{q}_i q ^ k)_{V-A}
		(\overline{q}_j c)_{V-A} \,,
	\end{eqnarray}
	where $c_\pm$ are the Wilson coefficients. 
 In this work, we use $i,j,k,l \in \{1,2,3\} $ as flavor indices with 
 $( q_1,q_2,q_3)=(u,d,s)$ which forms the fundamental representation of $SU(3)_F$.  
The
${\cal H} (\overline {\bf 6})$ and ${\cal H}({\bf 15})$ are tensors of $SU(3)_F$ whose non-zero entries are given by 
\begin{eqnarray}
&&{\cal H} ({\bf 15} ) ^{ 13  }_2  = {\cal H}(\overline{{\bf 6}})_{ 22  }
= V_{ud} V_{cs}^* \,, \nonumber\\
&&{\cal H} ({\bf 15} ) ^{ 12  }_2 
=- {\cal H} ({\bf 15} ) ^{ 13  }_3 
 = {\cal H}(\overline{{\bf 6}})_{ 23  }
=   V_{ud} V_{cd}^*\,,  \nonumber\\
&&{\cal H} ({\bf 15} ) ^{ 12  }_3  = {\cal H}(\overline{{\bf 6}})_{ 33  }
= V_{us} V_{cd}^*\,, 
\end{eqnarray}
while the other nonvanishing elements are obtained by using 
${\cal H}({\bf 15})^{ij}_k =
{\cal H}({\bf 15})^{ji}_k$  and ${\cal H}(\overline{{\bf 6}})_{ij}  ={\cal H}(\overline{{\bf 6}})_{ji} .$ 
The symmetric structures indicate  $\overline{q}_i$ and $\overline{q}_j$  in Eq.~\eqref{3}
are color-symmetric for the  term originated from ${\cal H}({\bf 15})$ whereas antisymmetric from ${\cal H}(\overline{{\bf 6}})$.  
The same also applies to $q^k$ and $c$. 
Here we have omitted \( {\cal H}({{\bf 3}}) = ( V_{ub}V^*_{cb} ,0,0 ) \), which has a minimal impact on CP-even quantities.

Since the decay amplitudes must remain invariant under the $SU(3)_F$ transformation in the symmetric limit, they should be  $SU(3)_F$ singlets.
Accordingly, the flavor indices are fully contracted. For different ways of contraction, we assign an undetermined parameter, respectively. Then 
the decay amplitudes are  decomposed into  several invariant amplitudes  as~\cite{Xing:2023dni,Asymmetries} 
\begin{widetext}
\begin{eqnarray}\label{eq22}
F(T_{c\overline{3}} \to {\bf B} P ) &=&
	 \tilde{f} ^a (P^\dagger) ^l_l   	 {\cal H}(\overline{{\bf 6}})_{ij}T^{ik}_c({\bf B}^\dagger )_k^j+ 
	\tilde{f}^b {\cal H}(\overline{{\bf 6}})_{ij}T_c^{ik}({\bf B}^\dagger )_k^l (P^\dagger)_l^j+\tilde{f}^c {\cal H}(\overline{{\bf 6}})_{ij}T^{ik}_c (P^\dagger)_k^l({\bf B}^\dagger)_l^j \nonumber\\
	&&	
+ \tilde{f}^d {\cal H}(\overline{{\bf 6}})_{ij}({\bf B}^\dagger)_k^i (P^\dagger)_l^j T_c^{kl}	+\tilde{f} ^e ({\bf B}^\dagger )^j_i {\cal H}({\bf 15})^{ ik }_l (P^\dagger)^l_k (T_{c \bar 3})_j
	\,, \nonumber\\
	G(T_{c\overline{3}} \to {\bf B} P ) &=& F(\tilde{f}^{x} \rightarrow  \tilde{g}^{x}  )\,, 
\end{eqnarray}
\end{widetext} 
with $T^{ij}_c = \epsilon^{ijk}(T_{c \bar 3})_k$  and  \(x \in \{ a, b, c, d, e \}\). 
These amplitudes will be expressed as \(\tilde{f}^{x} = f ^x \exp (i \delta_f^x)\) and \(\tilde{g}^{x} = g ^x \exp (i \delta_g^x)\), where \(f^x\) and \(g^x\) are strictly positive.


 Considering only the flavor structure, there are five different ways of contracting the  $SU(3)_F$ indices  for ${\cal H}({\bf 15})$: $(T_{c\bar{3}})_i {\cal H}({\bf 15})^{\{ik\}}_j({\bf B}^\dagger)^{ j}_kP^l_l$, 
$(T_{c\bar{3}})_i {\cal H}({\bf 15}) ^{\{ik\}}_j({\bf B}^\dagger)^{ l}_k $\,$ P^j_l$,
$(T_{c\bar{3}})_i {\cal H}({\bf 15})^{\{ik\}}_j ({\bf B}^\dagger)^{ j}_l$ $P^l_k$, $(T_{c\bar{3}})_i {\cal H}({\bf 15})^{\{jk\}}_l({\bf B}^\dagger)^{ l}_jP^i_k$ and 
$(T_{c\bar{3}})_i $ ${\cal H}({\bf15})^{\{jk\}}_l({\bf B}^\dagger)^{ i}_jP^l_k$. However, after taking into account of  that the color indices of the quarks originated from ${\cal H}({\bf 15})$~(baryons) must be (anti)symmetric, these five terms  can be reduced into one proportional to
$({\bf B}^\dagger )^j_i {\cal H}({\bf 15})^{\{ik\}}_l (P^\dagger)^l_k (T_{c \bar 3})_j$.   This is a remarkable result of the KPW theorem~\cite{Korner:1970xq,Groote:2021pxt} for hyperon decays when applied to charmed baryon decays. This reduction of the number of decay amplitudes enable us to use available data to completely determine the flavor $SU(3)_F$ decay amplitudes.

In the following,  we elucidate the configuration of the $SU(3)_F$ global fit.
Given that both 
$F$ and $G$ encompass five complex amplitudes each, by omitting one unphysical overall phase shift, say $\delta_{f}^b$, we are left with a total of 19 parameters.
If one does not consider decays involving $\eta$ and  $\eta'$, one can further neglect  $\tilde{f}^a$ and $\tilde{g}^a$. In that case there are only 15 parameters to work with. On the other hand, there are in total 29 (23 without $\eta$ and $\eta'$ data points) experimental data points~\cite{BelleDATA,ParticleDataGroup:2022pth}. The $SU(3)_F$ invariant amplitudes can therefore be completely determined from a global fit. The experimental data are listed in TABLE~\ref{EXP}. 
Note that had one kept the sub-leading  terms in  ${\cal H}({\bf 15})$, a global analysis becomes impossible at present.

We determine the best fit values for the decay amplitudes $\tilde f ^x$ and $\tilde g^x  $ by minimizing the $\chi^2$ function defined as 
\begin{equation}
\chi^2 (\tilde{f}^x,\tilde{g}^x) = 
\sum_{\text{exp}}\left( 
\frac{ O_{\text{th}}(\tilde{f}^x,\tilde{g}^x  ) - O_{\text{exp}}
}{\sigma_{\text{exp}}} 
\right) ^2 \,,
\end{equation}
where $O_{\text{th}}$ is the theoretical value of an observable, and  
$O_{\text{exp}}$ is the experimental value with the standard deviation of $\sigma_{\text{exp}}$. 
In conducting the global fit, we incorporate all of the experimental branching ratios and asymmetry parameters, \( \alpha_i \), available to date. 
For the decays of \( \Lambda_c^+ \) and  \( \Xi_c^+ \),  and ${\cal B}(\Xi_c^0 \to \Xi^- \pi^+)=(1.43\pm0.32)\%$, we rely on the absolute branching ratios documented in the Particle Data Group~\cite{ParticleDataGroup:2022pth}. While for the others, the reported ratios of \( {\cal R}_X  : = {\cal B}(\Xi_c^0\to X ) /{\cal B}(\Xi_c^0\to \Xi^-\pi^+) \) from Belle are utilized~\cite{Belle1,Belle6}. 
Employing \( {\cal R}_X \) as opposed to \( {\cal B}(\Xi_c^0\to X ) \) is crucial, as the former is what has been actually measured. 
Although measurements exist for  \( \beta_i \), their associated uncertainties are substantial~\cite{BESIII:2019odb}, making them insignificant to \( \chi^2 \). Consequently, they will not be incorporated into the fit.

The resultant best fit values of the decay amplitude parameters and the error bars are given  as follows:
\begin{widetext}
	{\small 
\begin{eqnarray}
&&f^x = 3.60(70 ), 3.64(1.20) , 3.84 ( 0.18), 1.25 ( 1.24), 2.19 ( 2.52) , ~g^x = 
12.21 (3.34), 28.05 ( 1.18),2.76 ( 1.72),5.23 ( 1.55), 6.49 ( 5.35)\,, \nonumber \\
&&\delta_f^x = 1.66(31),0,-2.20 ( 39), -0.57 ( 31), -0.58 ( 50) \,,
 ~~~\delta_g^x =  
-1.77(34), 2.60 ( 0.37),  2.03 ( 0.43),  2.39 ( 0.74),  1.98 ( 1.03)\,,
\end{eqnarray}} 
\end{widetext} 
in the order  of $x = a,b,c,d,e$, respectively,  and 
both \(f^x\) and \(g^x\) are in units of \(10^{-2}G_F \text{ GeV}^2\).  
We note that without measurement results for $\beta_i$, $\chi^2$  suffers from a $Z_2$ ambiguity of $(\delta_f^x ,\delta_g^x)\rightarrow(- \delta_f^x ,- \delta_g^x)$.
In addition, at the $SU(3)_F$ limit, $\kappa$ are the same  for all channels
resulting in an 
additional ambiguity  of 
$(f^x,\delta_f^x)\leftrightarrow (\kappa g^x, - \delta_g^x)$ 
without measured  $\gamma_i$ as input. 
To break the degeneracies,  we fix it by using 
$\beta(\Lambda_c^+ \to \Xi^0 K^+)>0$  and $\gamma(\Lambda_c^+ \to \Lambda^0 \pi^+)<0$
from the experiment~\cite{BESIII:2019odb,BESIII10}. 

 The  fitting values for ${\cal B}$, $\alpha$, $\beta$ and $\gamma$ are collected  in TABLE~\ref{EXP} for the observed decay modes. 
The presence of empty cells signifies that either the corresponding \( \alpha_{\text{exp}} \) values are absent, or the theoretical framework does not impose any constraints on those particular quantities. 
Asterisks are used to denote the number of standard deviations by which the observed values deviate from the theoretical central values.
Predictions for the unobserved decays with 148 decay observables  are collected in TABLE~\ref{CF} for the future experiment   verification.




Note that the phases of  $\delta_{f,g}^c $ are  sizable which give phase shift to the decay amplitude of  $F(\Lambda_c^+ \to \Xi^0 K^+) = -\tilde{f}^c$ and $G(\Lambda_c^+ \to \Xi^0 K^+) = -\tilde{g}^c$. 
In particular, we have $\delta_P -\delta_S =- 2.06 \pm0.50 $, which is consistent with the experimental finding of $-1.55\pm0.25$. 
As $\alpha \propto \cos (\delta_P - \delta_S) $,
this is crucial in obtaining a small value of $\alpha(\Lambda_c^+ \to \Xi^0 K^+)=  -0.15\pm 0.14$ to be consistent with the  BESIII measurement.
If we remove $\alpha(\Lambda^{+}_c\rightarrow\Xi^0K^+)=0.01\pm0.16$ as an input in the $SU(3)_F$ fit, then $\delta_P -\delta_S = -2.82 \pm 0.51$. The new BESIII data pulled the strong phase shift away from $-\pi$ which corresponds to no strong phase shift. Therefore, the data for $\alpha(\Lambda^{+}_c\rightarrow\Xi^0K^+)$ is crucial for reflecting the phase shift in $\Lambda^{+}_c \to \Xi^0 K^+$.  
In  addition, in contrast to the previous $SU(3)_F$ literature with $\alpha( \Lambda_c^+ \to \Sigma^0 K^+) \approx -1 $~\cite{Asymmetries,Huang:2021aqu,Zhong:2022exp,Xing:2023dni}, we find that $\alpha( \Lambda_c^+ \to \Sigma^0 K^+) = -0.52 \pm 0.10 $ which is consistent with the current experimental data. 

The establishment of  sizable strong phases
makes the study of CP violations 
possible~\cite{Donoghue:1986hh,Lenz:2020awd}  when combined with theoretical calculation for ${\cal H}( {\bf 3})$ contributions.
The direct CP asymmetries are expected to be at the size of $10^{-3}$ with the details 
given elsewhere.

	There are several direct relations appear when the color symmetry is considered. 
	In particular,  $\Gamma^{\Lambda_c^+}_{\Sigma^+ K_S} = \Gamma^{\Lambda_c^+}_{\Sigma^0 K^+}$  is well  satisfied by the experimental data~\cite{ParticleDataGroup:2022pth}, which
	partly justifies our approach in this work. 
An important new relation is 
	\begin{eqnarray}\label{6}  
	&& 	\frac{\tau_{\Lambda_c^+}}{\tau_{\Xi_c^0}}  {\cal B}( \Xi_c^0 \to \Xi^- \pi^+)  
		=
		{\cal B}( \Lambda_c^+ \to \Sigma^0 \pi^+)  \\
&&~~~+  
		3 {\cal B}( \Lambda_c^+ \to \Lambda \pi^+) - \left| \frac{V_{ud}}{V_{cd}} \right|^2 
		{\cal B}(\Lambda_c^+ \to n \pi^+ )  \,.\nonumber
	\end{eqnarray}
	By plugging the measured data at  BESIII  for $\Lambda_c^+$ decays on the right, we obtain
	${\cal B}(\Xi_c^0 \to \Xi^- \pi^+) = (2.96\pm0.31)\%$, which differs significantly to 
	$(1.43\pm 0.32 )\%$
	from PDG~\cite{ParticleDataGroup:2022pth}. 
	On the other hand,  with the relation of $\Gamma^{\Lambda_c^+} _{ \Xi^0 K^+} = \Gamma^{\Xi_c^0}_{\Sigma^+K^-}$, the experimental values of ${\cal B}(\Lambda_c^+ \to \Xi^0 K^+)$ 
	and  ${\cal R}^{\text{exp}}_{\Sigma ^+  K^-}  = 0.123 \pm 0.012$~\cite{Belle6}
	collectively	lead  to 
	${\cal B}(\Xi_c^0 \to \Xi^- \pi^+)= (3.37\pm 0.52)\%$, echoing with the discussion above. 
Intriguingly, the current algebra approach, exemplary in the $\Lambda_c^+$ sector, predicts ${\cal B}(\Xi_c^0 \to \Xi^- \pi^+) = 6.47\%$~\cite{Charmed-Cheng}.

The study of semi-leptonic decays might offer further insights on ${\cal B}(\Xi_c^0 \to \Xi^- \pi^+)$. Using the experimental ratio ${\cal R}_{ \Xi^- e^+\nu_e}^{\text{exp}}  = 0.730 \pm 0.044$~\cite{Belle:2021crz} alongside ${\cal B}(\Xi_c^0 \to \Xi^-\pi^+) = (2.72\pm 0.09 )\%$ from Table~\ref{EXP}, we derive ${\cal B}(\Xi_c^0 \to \Xi^- e^+\nu_e) = (1.98\pm 0.12)\%$. This aligns with ${\cal B}(\Xi_c^0 \to \Xi^- e^+\nu_e) = (2.38\pm 0.44)\%$ from lattice QCD~\cite{Zhang:2021oja} and ${\cal B}(\Xi_c^0 \to \Xi^- e^+\nu_e) = (2.17\pm 0.20)\%$ under the exact $SU(3)_F$ symmetry~\cite{He:2021qnc}. 
Conversely, ${\cal B}(\Xi_c^0 \to \Xi^-\pi^+) = (1.43\pm 0.32)\%$ implies ${\cal B}(\Xi_c^0 \to \Xi^- e^+\nu_e) = (1.04\pm 0.24)\%$, a deviation from the lattice QCD by $2.6\sigma$. 
We note that the latest preliminary result of lattice QCD indicates also a larger ${\cal B}(\Xi_c^0 \to \Xi^- e^+ \nu_e ) $~\cite{Farrell:2023vnm}. 


The above analysis indicates that the current experimental value for \({\cal B}(\Xi_c^0 \to \Xi^- \pi^+)\) might have underestimated its true value which needs to be scrutinized more carefully further. 
The global fit here yields a $\chi^2$ per degree of freedom($\chi^2/d.o.f$) value of $3.7$. This cannot be considered to be a good fit. The largest contributions to $\chi^2$ come from the experimental ratio of ${\cal R}^{\text{exp}}_{\Xi^- K^+}= 0.275\pm 0.057$ and ${\cal B} (\Xi_c^0 \to \Xi^- \pi^+) = (1.43\pm 0.32)\%$. 
If one removes these two data points, $\chi^2/d.o.f$ is reduced to $ 1.5$ which is a much better fit. 
	If we use 
	the original data of 
	${\cal B}(\Xi_c^0\to \Xi^-\pi^+) = (1.80 \pm 0.52)\%$~\cite{Belle:2018kzz}
	from  Belle, the overall $\chi^2/d.o.f.$ reduces to  1.9 indicating a better overall fit.  
Meanwhile, in these cases the predictions for other quantities do not change much. 
	We
	therefore
	 would like to emphasize that our analysis hints that ${\cal B}(\Xi_c^0\to \Xi^-\pi^+)=(1.43\pm0.32)\%$ is inconsistent with the direct relation of the non-leptonic decays and the indirect relation from the semi-leptonic decays.
	 

Should forthcoming experimental results confirm the diminished magnitude of ${\cal B}(\Xi_c^0 \to \Xi^- \pi^+)$,  the presence of a substantial gluon component within the $\Xi_c^0$ should be considered. In such a scenario, a rigorous examination of the sub-leading terms from ${\cal H}({\bf 15})$ would become imperative in theoretical discussions.

In this analysis, we have assumed \(SU(3)_F\) symmetry to be exact and the applicability of the KPW theorem to the processes under consideration. Possible corrections from \(SU(3)_F\) symmetry breaking effects due to quark mass differences and, also, KPW theorem breaking effects due to baryon states containing gluon Fock states, have been neglected. When the experimental data become more comprehensive, it is advisable to consider these effects to achieve greater theoretical precision. At the current stage, treating the \(SU(3)_F\) symmetry and the KPW theorem as exact offers valuable guidance. Future experimental data will provide further insights.

Finally we would also like to point out that the establishment of strong phases in the decay amplitudes have far reaching implications for opportunities of finding CP violations in   charmed baryon decays. One expects to have non-zero CP violating rate asymmetries for charmed baryon decays. Experimental searches should be carried out. We will present related theoretical studies elsewhere.



\renewcommand{\arraystretch}{1.4}

\begin{widetext}
	
\begin{table}[t]
	\caption{\label{brresult}
Predictions from the $SU(3)_F$ global fit for the observed decays. 
The experimental uncertainties are combined quadratically, and
the numbers in the parentheses are the uncertainties counting backward in digits, for example, $1.59(8) = 1.59\pm 0.08$.
The empty cells in the table indicate either $\alpha_{\text{exp}}  $ are missing or the theory imposes no constraint on the quantities.  
Asterisks denote the numbers of standard deviations against the theory. 
	}
	\label{EXP}
	\begin{tabular}{l|cc|cccccc}
		\hline
		\hline
		Channels &${\cal B}_{\text{exp}}(\%)$ &$\alpha_{\text{exp}}$&${\cal B}(\%)$&$\alpha$&$\beta$ &$\gamma$ \\
		\hline
$\Lambda_{c}^{+} \to p K_S $&$ 1.59 ( 8 )$&$^* 0.18 ( 45 )$&$ 1.55 ( 7 )$&$ -0.40 ( 49 )$&$ 0.32 ( 29 )$&$ -0.86 ( 19 )$\\
$\Lambda_{c}^{+} \to \Lambda^{0} \pi^{+} $&$ 1.30 ( 6 )$&$ -0.755 ( 6 )$&$ 1.29 ( 5 )$&$ -0.75 ( 1 )$&$ -0.13 ( 19 )$&$ -0.64 ( 4 )$\\
$\Lambda_{c}^{+} \to \Sigma^{0} \pi^{+} $&$ 1.27 ( 6 )$&$ -0.466 ( 18 )$&$ 1.27 ( 5 )$&$ -0.47 ( 2 )$&$ 0.88 ( 2 )$&$ -0.05 ( 27 )$\\
$\Lambda_{c}^{+} \to \Sigma^{+} \pi^{0} $&$ 1.25 ( 10 )$&$ -0.48 ( 3 )$&$ 1.27 ( 5 )$&$ -0.47 ( 2 )$&$ 0.88 ( 2 )$&$ -0.05 ( 27 )$\\
$\Lambda_{c}^{+} \to \Xi^{0} K^{+} $&$ ^{**}0.55 ( 7 )$&$ 0.01 ( 16 )$&$ 0.40 ( 3 )$&$ -0.15 ( 14 )$&$ -0.29 ( 22 )$&$ 0.94 ( 7 )$\\
$\Lambda_{c}^{+} \to \Lambda^{0} K^{+} $&$ 0.064 ( 3 )$&$ -0.585 ( 52 )$&$ 0.063 ( 3 )$&$ -0.56 ( 5 )$&$ 0.82 ( 5 )$&$ 0.10 ( 27 )$\\
$\Lambda_{c}^{+} \to \Sigma^{0} K^{+} $&$ 0.0382 ( 25 )$&$ -0.54 ( 20 )$&$ 0.0365 ( 21 )$&$ -0.52 ( 10 )$&$ 0.48 ( 24 )$&$ -0.71 ( 17 )$\\
$\Lambda_{c}^{+} \to n \pi^{+} $&$ 0.066 ( 13 )$&&$ 0.067 ( 8 )$&$ -0.78 ( 12 )$&$ -0.63 ( 15 )$&$ -0.04 ( 20 )$\\
$\Lambda_{c}^{+} \to \Sigma^{+} K_S $&$ 0.048 ( 14 )$&&$ 0.036 ( 2 )$&$ -0.52 ( 10 )$&$ 0.48 ( 24 )$&$ -0.71 ( 17 )$\\
$\Lambda_{c}^{+} \to p \pi^{0} $&$ < 0.008 $&&$ 0.02 ( 1 )$& &$- 0.82 ( 32 )$&$ 0.57 ( 48 )$\\
$\Lambda_{c}^{+} \to \Sigma^{+} \eta $&$ 0.32 ( 4 )$&$ -0.99 ( 6 )$&$ 0.32 ( 4 )$&$ -0.93 ( 4 )$&$ -0.32 ( 16 )$&$ -0.16 ( 23 )$\\
$\Lambda_{c}^{+} \to p \eta $&$ 0.142 ( 12 )$&&$ 0.145 ( 26 )$&$ -0.42 ( 61 )$&$ 0.64 ( 40 )$&$ -0.65 ( 20 )$\\
$\Lambda_{c}^{+} \to \Sigma^{+} \eta' $&$ 0.437 ( 84 )$&$ -0.46 ( 7 )$&$ 0.420 ( 70 )$&$ -0.44 ( 25 )$&$ 0.86 ( 6 )$&$ 0.25 ( 35 )$\\
$\Lambda_{c}^{+} \to p \eta' $&$ 0.0484 ( 91 )$&&$ 0.0520 ( 114 )$&$ -0.59 ( 9 )$&$ 0.76 ( 14 )$&$ -0.26 ( 33 )$\\
$\Xi_{c}^{+} \to \Xi^{0} \pi^{+} $&$ 1.6 ( 8 )$&&$ 0.90 ( 16 )$&$ -0.94 ( 6 )$&$ 0.32 ( 21 )$&$ -0.07 ( 20 )$\\
$\Xi_{c}^{0} \to \Xi^{-} \pi^{+} $&$ ^{****}1.43 ( 32 )$&$ ^{*}-0.64 ( 5 )$&$ 2.72 ( 9 )$&$ -0.71 ( 3 )$&$ 0.36 ( 20 )$&$ -0.60 ( 12 )$\\
\hline
Channels &${\cal R}^{\text{exp}}_{X}$ &$\alpha_{\text{exp}}$&${\cal R}_{X}$&$\alpha$&$\beta$ &$\gamma$ \\
\hline
$\Xi_{c}^{0} \to \Lambda^{0} K_S $&$ 0.225 ( 13 )$&&$ 0.233 ( 9 )$&$ -0.47 ( 29 )$&$ 0.66 ( 20 )$&$ -0.58 ( 21 )$\\
$\Xi_{c}^{0} \to \Xi^{-} K^{+} $&$^{**} 0.0275 ( 57 )$&&$ 0.0410 ( 4 )$&$ -0.75 ( 4 )$&$ 0.38 ( 20 )$&$ -0.55 ( 13 )$\\
$\Xi_{c}^{0} \to \Sigma^{0} K_S $&$ 0.038 ( 7 )$&&$ 0.038 ( 7 )$&$ -0.07 ( 117 )$&$ -0.83 ( 28 )$&$ 0.55 ( 41 )$\\
$\Xi_{c}^{0} \to \Sigma^{+} K^{-} $&$ 0.123 ( 12 )$&&$ 0.132 ( 11 )$&$ -0.21 ( 18 )$&$- 0.39 ( 29 )$&$ 0.90 ( 13 )$\\
		\hline
		\hline	
	\end{tabular}
\end{table}

\begin{table}[t]
	\footnotesize
	\caption{Legend as in TABLE~\ref{EXP} but for unobserved decays.
	}
	\label{CF}
	\begin{tabular}{lcccc||lcccc}
		\hline
		\hline
		Channels &${\cal B}(10^{-3})$&$\alpha$&$\beta$ &$\gamma$&Channels&${\cal B}(10^{-4})$&$\alpha$&$\beta$ &$\gamma$ \\
		\hline
		$\Lambda_{c}^{+} \to p K_L $&$ 15.20 ( 67 )$&$ -0.40 ( 44 )$&$ 0.33 ( 27 )$&$ -0.86 ( 17 )$& 
		$\Lambda_{c}^{+} \to n K^{+} $&$ 0.13 ( 2 )$&$ -0.90 ( 6 )$&$ 0.31 ( 20 )$&$ -0.32 ( 2 )$\\
		\hline
		$\Xi_{c}^{+} \to \Sigma^{+} K_S $&$ 0.59 ( 49 )$ &&&&
		$\Xi_{c}^{+} \to \Lambda^{0} \pi^{+} $&$ 3.24 ( 90 )$&$ 0.29 ( 29 )$&$ -0.47 ( 30 )$&$ -0.83 ( 13 )$\\
		$\Xi_{c}^{+} \to p K_{S/L} $&$ 1.90 ( 15 ) $&$ -0.37 ( 7 )$&$ 0.35 ( 18 )$&$ -0.86 ( 8 )$&
		$\Xi_{c}^{+} \to n \pi^{+} $&$ 0.34( 4 )$&$ -0.27 ( 23 )$&$ -0.51 ( 35 )$&$ 0.81 ( 26 )$\\
		$\Xi_{c}^{+} \to \Sigma^{+} \pi^{0} $&$ 2.12  ( 14 )$&$ -0.49 ( 49 )$&$ 0.83 ( 26 )$&$ -0.26 ( 27 )$&
		$\Xi_{c}^{+} \to \Sigma^{0} K^{+} $&$ 1.17 ( 4 )$&$ -0.68 ( 3 )$&$0.35 ( 19 )$&$ -0.65 ( 26 )$\\
		$\Xi_{c}^{+} \to \Sigma^{+} \eta $&$ 0.70( 33 )$&$ -0.80 ( 69 )$&$ -0.41 ( 77 )$&$ -0.43 ( 104 )$&
		$\Xi_{c}^{+} \to p \pi^{0} $&$ 0.17 ( 2 )$&$ -0.27 ( 23 )$&$ -0.51 ( 35 )$&$ 0.81 ( 26 )$\\
		$\Xi_{c}^{+} \to \Sigma^{+} \eta' $&$ 1.13 ( 24 )$&$ -0.44 ( 30 )$&$ 0.88 ( 23 )$&$ -0.19 ( 42 )$&
		$\Xi_{c}^{+} \to p \eta $&$ 1.72 ( 37 )$&$ -0.41 ( 7 )$&$ 0.67 ( 15 )$&$ -0.62 ( 26 )$\\
		$\Xi_{c}^{+} \to \Sigma^{0} \pi^{+} $&$3.04 ( 10 )$&$ -0.59 ( 3 )$&$ 0.75 ( 7 )$&$ -0.29 ( 22 )$&
		$\Xi_{c}^{+} \to p \eta' $&$ 0.94 ( 18 )$&$ -0.53 ( 5 )$&$ 0.73 ( 18 )$&$ -0.43 ( 26 )$\\
		$\Xi_{c}^{+} \to \Xi^{0} K^{+} $&$ 1.04 ( 14 )$&$ -0.73 ( 12 )$&$ -0.59 ( 14 )$&$ 0.35 ( 17 )$&
		$\Xi_{c}^{+} \to \Lambda^{0} K^{+} $&$ 0.37 ( 4 )$&$ -0.44 ( 12 )$&$ 0.63 ( 21 )$&$ 0.65 ( 26 )$\\
		\hline
		$\Xi_{c}^{0} \to \Sigma^{0} K_L $&$ 0.97 ( 17 )$&&$ -0.53 ( 39 )$&$ 0.84 ( 28 )$&
		$\Xi_{c}^{0} \to p K^{-} $&$ 1.96 ( 19 )$&$ -0.26 ( 22 )$&$ -0.50 ( 34 )$&$ 0.83 ( 20 )$\\
		$\Xi_{c}^{0} \to \Xi^{0} \pi^{0} $&$ 7.10 ( 41 )$&$ -0.49 ( 9 )$&$0.46 ( 23 )$&$ -0.74 ( 15 )$&
		$\Xi_{c}^{0} \to n K_{S/L} $&$ 7.10 ( 62 )$&$ -0.44 ( 3 )$&$ 0.83 ( 8 )$&$ -0.36 ( 23 )$\\
		$\Xi_{c}^{0} \to \Xi^{0} \eta $&$ 2.94 ( 97 )$&$ 0.04 ( 22 )$&$ 0.83 ( 13 )$&$ 0.55 ( 21 )$&
		$\Xi_{c}^{0} \to \Lambda^{0} \pi^{0} $&$ 0.89 ( 17 )$&$ -0.32 ( 50 )$&$ -0.40 ( 31 )$&$ -0.86 ( 24 )$\\
		$\Xi_{c}^{0} \to \Xi^{0} \eta' $&$ 5.66 ( 93 )$&$ -0.58 ( 15 )$&$ 0.74 ( 6 )$&$ 0.34 ( 25 )$&
		$\Xi_{c}^{0} \to n \pi^{0} $&$ 0.06 ( 1 )$&$ -0.27 ( 23 )$&$ -0.51 ( 35 )$&$ 0.81 ( 26 )$\\
		$\Xi_{c}^{0} \to \Lambda^{0} K_L $&$ 7.07 ( 24 )$&$ -0.47 ( 24 )$&$ 0.71 ( 17 )$&$ -0.53 ( 21 )$&
		$\Xi_{c}^{0} \to \Lambda^{0} \eta $&$ 4.31 ( 1.10 )$&$ -0.02 ( 52 )$&$ 0.12 ( 30 )$&$ -0.99 ( 2 )$\\
		$\Xi_{c}^{0} \to \Sigma^{+} \pi^{-} $&$ 0.21 ( 2 )$&$ -0.22 ( 19 )$&$ -0.41 ( 30 )$&$ 0.88 ( 14 )$&
		$\Xi_{c}^{0} \to \Lambda^{0} \eta' $&$ 6.83 ( 1.32 )$&$ -0.67 ( 6 )$&$ 0.74 ( 8 )$&$ -0.09 ( 26 )$\\
		$\Xi_{c}^{0} \to \Sigma^{0} \pi^{0} $&$ 0.34 ( 3 )$&$ -0.33 ( 48 )$&$- 0.38 ( 27 )$&$ -0.87 ( 23 )$&
		$\Xi_{c}^{0} \to \Sigma^{-} K^{+} $&$ 0.78 ( 3 )$&$ -0.68 ( 3 )$&$ 0.35 ( 19 )$&$ -0.65 ( 26 )$\\
		$\Xi_{c}^{0} \to \Sigma^{0} \eta $&$ 0.12 ( 5 )$&$ -0.80 ( 69 )$&$- 0.41 ( 77 )$&$ -0.43 ( 104 )$&
		$\Xi_{c}^{0} \to p \pi^{-} $&$ 0.11 ( 1 ) $&$ -0.27 ( 23 )$&$ -0.51 ( 35 )$&$ 0.81 ( 26 )$\\
		$\Xi_{c}^{0} \to \Sigma^{0} \eta' $&$ 0.19 ( 4 )$&$ -0.44 ( 30 )$&$ 0.88 ( 23 )$&$ -0.19 ( 42 )$&
		$\Xi_{c}^{0} \to n \eta' $&$ 0.31 ( 6 ) $&$ -0.53 ( 5 )$&$ 0.73 ( 18 )$&$ -0.43 ( 26 )$\\
		$\Xi_{c}^{0} \to \Sigma^{-} \pi^{+} $&$ 1.83 ( 6 )$&$ -0.65 ( 3 )$&$ 0.33 ( 18 )$&$ -0.69 ( 9 )$&
		$\Xi_{c}^{0} \to n \eta $&$ 0.57 ( 12)$&$ -0.41 ( 7 )$&$ 0.67 ( 15 )$&$ -0.62 ( 26 )$\\
		$\Xi_{c}^{0} \to \Xi^{0} K_{S/L} $&$ 0.43 ( 2 )$&$ -0.47 ( 2 )$&$ 0.88 ( 1 )$&$ 0.06 ( 26 )$&\\
		\hline
	\end{tabular}
\end{table}

\end{widetext}

\begin{acknowledgments}

This work is supported in part by 
the National Key Research and Development Program of China under Grant No. 2020YFC2201501, by
the Fundamental Research Funds for the Central Universities, by National Natural Science Foundation of P.R. China (No.12090064, 11735010, 11985149 and 12205063).  XGH is also supported in part by MOST 109–2112-M-002–017-MY3.

\end{acknowledgments}


\begin{thebibliography}{9}
\bibitem{idk}
	We note that two solutions for \(\delta_P - \delta_S\) are related by \(\pi\) in modulus. This is because the phase in experiments is extracted using \(\tan(\delta_P - \delta_S) = \beta/\alpha\)~\cite{BESIII10}.
\end{thebibliography}
\end{document}